\documentclass[useAMS,usenatbib]{mnras}
\usepackage{graphicx} 
\usepackage{subcaption}
\usepackage{amsmath} 
\usepackage{xcolor}
\usepackage{float}
\usepackage{caption}
\usepackage{placeins}
\usepackage{lipsum}
\usepackage{multicol}
\usepackage{soul}
\usepackage{amssymb}
\usepackage{ragged2e}
\usepackage{threeparttable}

\def\refjnl#1{{\rm#1}}
\def\apjl{\refjnl{ApJ}}                
\def\apjs{\refjnl{ApJS}}               
\def\ao{\refjnl{Appl.~Opt.}}           
\def\aap{\refjnl{A\&A}}                

\def\be{\begin{equation}} 
\def\ee{\end{equation}}

\def\msun{{\Msun}}

\def\psiuv{\psi_{\rm UV}}

\def\gsim{\lower.5ex\hbox{\gtsima}} 
\def\lsim{\lower.5ex\hbox{\ltsima}} \def\gtsima{$\; \buildrel > \over 
\sim \;$} \def\ltsima{$\; \buildrel < \over \sim \;$} \def\prosima{$\; 
\buildrel \propto \over \sim \;$} \def\gsim{\lower.5ex\hbox{\gtsima}} 
\def\lsim{\lower.5ex\hbox{\ltsima}} 
\def\simgt{\lower.5ex\hbox{\gtsima}} 
\def\simlt{\lower.5ex\hbox{\ltsima}} 
\def\simpr{\lower.5ex\hbox{\prosima}} \def\la{\lsim} \def\ga{\gsim} 
  
 \def\gtsima{$\; \buildrel > \over \sim \;$} 
\def\ltsima{$\; \buildrel < \over \sim \;$} 
\def\gsim{\lower.5ex\hbox{\gtsima}} 
\def\lsim{\lower.5ex\hbox{\ltsima}} 
\def\simgt{\lower.5ex\hbox{\gtsima}} 
\def\simlt{\lower.5ex\hbox{\ltsima}} 
\def\simpr{\lower.5ex\hbox{\prosima}} 
\def\la{\lsim} 
\def\ga{\gsim}

\def\fescuv{f_{\rm esc}^{UV}}

\def\msun{\,{\rm \Msun}}

\def\E3{{\cal E}_{\rm g}^{III}}

\def\rhouv{\rho_{\rm UV}}
\def\msun{\rm M_\odot}

\def\Zsun{\rm Z_\odot}

\def\Msun{\rm M_\odot}
\def\myr{\rm Myr}

\def\zsun{\rm Z_\odot}
\def\M*{M_*}

\def\Z*{Z_*}
\def\L*{L_*}
\def\muv{\rm M_{UV}}

\newcommand\code[1]{\textsc{\MakeLowercase{#1}}}

\title[Bright galaxies at high-redshift]{Exploring the interplay between star formation efficiency and dust in regulating the UV luminosity of early systems in the JWST and ALMA era} 
\author[]{Giorgos P. Nikopoulos$^{1,2}$\thanks{georgios.nikopoulos@nbi.ku.dk} \& Pratika Dayal$^{3,4,5}$\thanks{pdayal@cita.utoronto.ca} \\ 
$^{1}$  Cosmic Dawn Center (DAWN), Copenhagen, Denmark \\
$^{2}$ Niels Bohr Institute, University of Copenhagen, Jagtvej 128, 2200 Copenhagen N, Denmark\\
$^{3}$ Canadian Institute for Theoretical Astrophysics, 60 St George St,
University of Toronto, Toronto, ON M5S 3H8, Canada\\
$^{4}$ David A. Dunlap Department of Astronomy and Astrophysics,
University of Toronto, 50 St George St, Toronto ON M5S 3H4,
Canada\\
$^{5}$ Department of Physics, 60 St George St, University of Toronto,
Toronto, ON M5S 3H8, Canada}

\begin{document} 
 
\date{} 

\maketitle

\begin{abstract}
James Webb Telescope (JWST) observations have unveiled numerous galaxy candidates between $z \sim 9 - 16.5$, hinting at an over-abundance of sources at the bright-end of the UV luminosity function (UV LF) at z $\gsim$ 11. Complementarily, the Atacama Large Millimetre Array (ALMA) has been yielding dust mass estimates at $z \sim 5 - 7$. In this work, we develop an analytic formalism baselined against ALMA results, jointly exploring the impact of bursty star formation and its associated dust enrichment, on the visibility of early galaxies, while also modelling sources scattered off the main sequence of star formation. We incorporate dust production in type II Supernovae, dust destruction, ejection, growth and sputtering. Our key results are: (i) explaining the UV LF at $z \sim 5 - 13$ requires an average star formation efficiency that evolves as $f_*(z) = 10^{0.13z-3.5}$, with a number of observations exceeding this main sequence by a factor of 10; (ii) The dust enrichment of early systems is driven by dust production in SNII ejecta, while growth and sputtering impact the dust mass by 60\% and 40\% respectively at $z \sim 7$; (iii)  galaxies at $z \gsim  9$ can retain significant dust, reaching average dust-to-stellar mass ratios of 0.19\% (0.14\%) at $z \sim 9$ ($z \sim 11$). Dust attenuation decreases with redshift as dust becomes increasingly dispersed within halos; (iv) observations by ALMA at $z \sim 5$ and 7 are not representative of the average population that makes up the UV LF; (v) assuming all stars to have formed instantaneously results in a high light-to-mass ratio. This naturally results in our model yielding a lower limit on the stellar mass contained in a halo, also under-predicting the observed stellar mass function.
\end{abstract}


\section{Introduction}

Galaxy formation in the first billion years and its large-scale effects remains a key frontier in completing our understanding of the cosmic timeline. Over the past few years, the James Webb Space Telescope (JWST) has opened a new window into this era by providing unprecedented views on galaxy formation as early as redshifts of $z \sim 9-16.5$ given its exquisite sensitivity \citep[e.g.][]{adams2022,naidu2022,atek2022,bradley2022, donnan2022, Castellano2023, Fujimoto2023a, Harikane2024}. Puzzlingly, these observations seem to indicate an over-abundance of bright systems at these early epochs, yielding an elevated bright end of the ultra-violet luminosity function (UV LF) at $z \gsim 11$ \citep[e.g.][]{Adams2023a,bouwens2022jwst,Bouwens2023b,Harikane2024b,Harikane2023b, Harikane2024,Leung2023,Mcleod2024, PerezGonzalez2023,Willot2023, donnan2024jwst, Whitler2025}. Although a number of works urge caution when using the LF at $z \gsim 12$ where the redshift and nature of the sources remains debated \citep{adams2022, naidu2022, popping2022}, a number of UV LF estimates at $z \gsim 12$ lie troublingly close to the maximum upper limit allowed by theory - this is when each halo is assigned a cosmological ratio of baryons to dark matter and all of these baryons form stars instantaneously \citep[e.g.][]{naidu2022,Mason2023,mauerhofer23dust}. 

Observational caveats notwithstanding, a number of theoretical approaches have been followed to explain these discordant observations including a decrease in dust attenuation with increasing redshift \citep{ferrara2022, mauerhofer23dust, Ferrara2024, nakazato2024}, an evolving initial mass function (IMF) of the stellar population \citep{Yung2024,Cueto2024,jeong2024, Mauerhofer2025}, stochastic or bursty star formation \citep{Mason2023, MirochaFurlanetto2023, Sun2023a, Sun2023b}, a high efficiency of gas conversion into stars \citep{Dekel2023, Renzini2023, SippleLidz2024,jeong2024} and even black hole contribution to the UV luminosity \citep{ono2018,pacucci2022}. 

The issue of dust enrichment and its impact on galaxy luminosity has become particularly pertinent given recent Atacama Large Millimetre Array (ALMA) observations that are yielding the first glimpses of dust within the first billion years. These include the ALMA Large Program to INvestigate C+ at Early Times \citep[ALPINE;][]{dessauges2020, bethermin2020} at $z \sim 4-6$ and the ALMA Reionization Epoch Bright Line Emission Survey at $z \sim 6.5-7.5$ \citep[REBELS;][]{bouwens2022, inami2022}. Tantalisingly, these observations yield 
extremely high dust-to-stellar-mass ratios (ranging between $0.012-3\%$) for star forming galaxies with stellar masses $M_* \sim 10^{8.3-10.5}\msun$, at $z \gsim 7$, in accord with a number of previous estimates \citep[e.g.][]{watson2015, laporte2017, hashimoto2019,bakx2020, reuter2020, bouwens2022}. They also suggest as much as 30\% of the total star formation rate density at $z \sim 7$ might be obscured by dust \citep{Algera2023}. This has led to a variety of approaches aimed at studying the key dust processes and dust enrichment of early galaxies, ranging from zoom-in simulations \citep[e.g.][]{zhukovska2016,hirashita2018, mckinnon2018} to cosmological simulations \citep{graziani2020,vogelsberger2020,DiCesare2023,Katz2023,Vijayan2024} to semi-analytic models \citep[e.g.][]{popping2017,vijayan2019, triani2020, dayal2022,mauerhofer23dust}. As might be expected, dust enrichment and its impact on the visibility of early systems crucially depends on the model assumptions of dust production and its spatial segregation as compared to star forming sites \citep[e.g.][]{inami2022}. 

In this work, we develop an analytic model that, for the first time, jointly explores the impact of two crucial components - bursty star formation rates (SFR) and the associated dust enrichment - on the visibility of early systems. Starting from the halo mass function, we use the faint-end of the observed UV LF at $z \sim 5-13$, to calibrate the star formation efficiency at a given redshift; crucially we also model sources that can be significantly scattered off the average main sequence of star formation. We then calculate the dust enrichment for a given stellar mass including all of the key dust processes using minimal free parameters - this is used to obtain the ``dust-attenuated" luminosity for each system. 

We describe this framework in detail in Sec. \ref{model}. We present the resulting evolution of the UV LF and UV luminosity density in Sec. \ref{results-uvlf} before discussing the dust mass-stellar mass relations and their redshift evolution in Sec. \ref{results-dustm-stellm} and the impact of dust on the UV-inferred SFRs in Sec. \ref{results-dust impact}. Finally, we show our prediction of the dust mass function in Sec. \ref{results-dustmassfn} before discussing the nature of obscured outliers at high-redshifts in Sec. \ref{discussion-datapoints} and conclude in Sec. \ref{conclusions-discussion}. We show the stellar mass functions resulting from our work, and their comparison with other theoretical works, in the appendix.

We adopt a $\Lambda$CDM model with dark energy, dark matter and baryonic density parameter values of $\Omega_{\Lambda}= 0.691$, $\Omega_{m}= 0.308$ and $\Omega_{b}= 0.049$, respectively, a Hubble constant $H_0=100\, h\,{\rm km}\,{\rm s}^{-1}\,{\rm Mpc}^{-1}$ with $h=0.67$, a spectral index of $n=0.96$ and a value of $\sigma_{8}=0.81$ \citep[][]{planck2016}. Throughout this work, we use a Salpeter IMF \citep{salpeter1955} between $1-100\msun$. Finally, we quote all quantities in comoving units, unless stated otherwise, and express all magnitudes in the standard AB system \citep{oke-gunn1983}.

\section{The theoretical model}
\label{model}
In this phenomenological model, we start by using the Sheth-Tormen halo mass function \citep[HMF;][]{sheth-tormen1999} at $z \sim 5-20$ in steps of $\Delta z =1$. In principle, any halo can have a maximum gas mass ($M_g$) that is given by the cosmological baryon-to-dark matter ratio such that $M_g = (\Omega_b/\Omega_m)M_h$ where $M_h$ is the halo mass. However, the gas content of low-mass halos in ionized regions can be suppressed as a result of the heating ultra-violet background (UVB) generated during the process of reionization \citep[e.g.][]{sobacchi2013b, hutter2021} such that \citep[e.g.][]{sobacchi2013b}
\be
M_g =   \bigg(\frac{\Omega_b}{\Omega_m}\bigg)~2^{- M_{\rm c} / M_h}~M_h.
\label{mg}
\ee
Here, $M_{\rm c}$ is a ``critical" halo mass which contains half of the cosmological baryon-to-dark matter ratio. We use a redshift-independent value of $M_{\rm c} = 10^{10}\msun$ in this work; we caution that this characteristic mass is degenerate with the volume filling fraction of ionized hydrogen in terms of its impact on the gas content of low-mass galaxies \citep[e.g.][]{choudhury2019}. A fraction of this total gas mass is assumed to form stars instantaneously with an efficiency $f_*$. We note that, in principle, only cold gas should be involved in the process of star formation; we assume our understanding of this fraction to be absorbed in $f_*$. This ``newly" formed stellar mass can be expressed as $M_* = f_* M_g$. In the spirit of maintaining simplicity, we assume every new stellar population has a fixed metallicity of $0.05 \Zsun$ and an age of $2 \, \myr$. Using these parameters (and our chosen IMF) with the population synthesis code {\small STARBURST99}, the specific intrinsic UV luminosity at $\lambda = 1500$\,\AA~ produced by this newly-formed stellar mass can be expressed as
\begin{equation}
L^{int}_{\rm 1500} = 10^{33.404} \bigg(\frac{M_{*}}{\Msun} \bigg) \,\, {\rm erg\, s}^{-1} {\rm \AA}^{-1}.
\label{lumnew}
\end{equation}
This UV luminosity can be used to infer an associated {\it total star formation rate} ($\psi$) as $ L_{\rm 1500} = \psi \kappa^{-1}$. To be able to compare with recent ALMA results \citep[e.g.][]{ferrara2022a}, we assume a standard fixed value of $\kappa = 4.45 \times 10^{-29}~ {\rm \msun~ yr^{-1}~ erg^{-1}~ s ~Hz}$. 

\subsection{Calculating the dust enrichment of early sources}
\label{model_dust}
 We make the reasonable assumption that Type II supernovae (SNII) are the primary dust factories at $z \sim 5$ given the long evolutionary timescales associated with other dust sources such as asymptotic giant branch (AGB) stars \citep[e.g.][]{dayal2010a, mancini2015, lesniewska2019, dayal2022}. We  calculate the total dust mass ($M_d$) produced by a newly formed stellar population as \citep[see also][]{dayal2022}
\be
M_{d} = M_d^{\rm pro} - M_d^{\rm ast} -  M_d^{\rm des} - M_d^{\rm eje} + M_d^{\rm gro} - M_d^{\rm sput},
\label{eq_dust}
\ee
where the terms on the right hand side account for dust production, the astration of dust into star formation, dust destruction in SNII shocks, dust ejection due to SNII powered winds, dust enhancement due to grain growth in the interstellar medium (ISM) and dust sputtering in the ISM, respectively. Since we model a single burst of star formation in any given halo, we ignore the astration term. While the processes of production, destruction and ejection can be treated as being instantaneous, grain growth and sputtering require accounting for their timescales, as now detailed. We also note that we assume perfect mixing of gas and dust as well as an equi-partition of gas into the hot and cold components for every process:
\begin{itemize}
\item The dust produced by SNII is calculated as $M_d^{\rm pro'} = y_d \nu M_*$ where we use a SNII dust yield of $y_d = 1.0\, \msun$. Further, the Salpeter IMF adopted in this work results in a SNII rate of $\nu = [53 \msun]^{-1}$. This yields a total dust mass of $M_d^{\rm pro'}= 0.0189 M_* ~ [\msun]$. However, only a fraction of this dust survives the reverse SN shock \citep[e.g.][]{bianchi2007}. Although the value remains debated, a general consensus is that dust survival increases with decreasing ISM density. We therefore use the results from \citet{bianchi2007} to calculate a dust survival fraction ($f_{rs}$) 
\begin{equation}
{\rm log}_{10}(f_{rs}) = -0.5~{\rm log}\bigg(\frac{\rho}{10^{-25} {\rm gm ~ cm^{-3}}}\bigg)-0.68,
\end{equation}
where $\rho$ is the average ISM density which is calculated as $\rho = 200 \rho_c(z)$ where $\rho_c(z)$ is the critical density at the given redshift. This results in a survival fraction that, whilst constant for all halos at a given $z$, decreases with increasing $z$ for any halo mass - for example, $f_{rs} \sim 0.19$ at $z\sim 5$ which decreases to $f_{rs} \sim 0.06$ by $z\sim 12$. The dust left after the reverse shock can then be calculated as
\be
M_d^{\rm pro} =  0.0189  f_{rs} M_*~ [\msun].
\label{dustprod}
\ee
As seen, at a given redshift, the dust mass left after the reverse shock is only a function of the star formation efficiency and halo mass.

\item The dust mass destroyed in SNII shocks is calculated as $M_d^{\rm des} = (1-X_c)f\,   \epsilon\,  \nu\, M_s\mathcal{D}\,M_*$. Here, ($1-X_c$) is the mass fraction of warm ISM where dust can be destroyed for which we use a {\it fiducial }value of $X_c=0.5$ based on high resolution zoom-in simulations of early galaxies \citep{Pallottini19}, $ f \sim 15\%$ is the SNII fraction contributing to such shocks \citep{debennassuti2014}, $\epsilon \sim 0.2$ is a reasonable value for the dust destruction efficiency in shocks \citep{mckee1989, seab1983} and $M_s = 6.8 \times 10^3 \, {\rm M_\odot}$ is the gas mass accelerated to $>100$ km ${\rm s^{-1}}$ by the SNII blast wave \citep{mckee1989, lisenfeld1998}. Finally, $\mathcal{D}$ is the dust-to-gas ratio, the upper limit to which is given by $M_d^{\rm pro}/M_g = \nu\,f_*\,f_{rs}$. Using the above noted values for the different parameters, we obtain
 \be
 M_d^{\rm des} = 0.0363 f_{rs} f_* M_*.
 \label{dustdes}
 \ee
As detailed at the end of this section, matching to the faint end of the UV LF at $z \sim 5 ~(10)$ requires $f_* \sim 1.4\times 10^{-3}~ (4.5\times 10^{-3})$. With these numbers, the net dust mass left after production and SNII-shock destruction at $z \sim 5$ is $\sim 3.7 \times 10^{-3} M_*$, in good agreement with estimates for dusty galaxies at this redshift \citep[e.g.][]{sommovigo2022b}.
 
 \item Assuming perfect mixing of dust and gas, the ejected dust mass is $M_d^{\rm eje} = \mathcal{D}  M_g^{\rm eje}$, where $M_g^{\rm eje}$ is the ejected gas mass. We calculate the ejected gas mass by comparing the SNII energy that couples to gas ($E_{SN}$) and the halo binding energy ($E_{bin}$) such that
 \be
 M_g^{\rm eje} = \frac{E_{SN}}{E_{bin}} M_g = M_g \frac{f_w E_{51} \nu M_*}{(M_g-M_*) v_c^2},
 \ee
where $f_w = 0.1$ is the fraction of SNII energy that couples to the gas \citep{dayal2022}, $E_{51}=10^{51}$ erg is the energy produced per SNII and $v_c$ is the halo circular velocity. 
This can be simplified to
\be
M_d^{\rm eje} = \mathcal{D}\frac{f_w}{(1-f_*)} \frac{v_s^2}{v_c^2}  M_* ,
\label{dustej}
\ee
where $v_s = (E_{51} \nu)^{1/2} = 973.97\,{\rm km ~ s^{-1}}$. Galaxies where the SNII energy exceeds the binding energy naturally lose all of their gas and dust contents. 

\item Grain growth occurs in the cold component of the ISM, where dust grains can accrete gas-phase metals. The grain growth rate is modelled as \citep{dwek1998} 
\be
\dot{M}_{d}^{gro} = (1 - \mathcal{D})\frac{X_c~M_{d}^\prime}{\tau_{acc}}.
\ee
Here the grain growth timescale is calculated as $\tau_{acc} = \tau_0 [{Z}/{Z_\odot}]^{-1}$ where $Z=0.05\zsun$ is the gas-phase metallicity and $M_{d}^\prime$ is the dust mass left after production, destruction and ejection. We assume grain growth acts over a 30 Myr timescale corresponding to the maximum age of SNII \citep{padovani1993} such that:
\be
M_d^{gro} = (1-0.0189\, f_{rs}\,f_*)\frac{3M_d^\prime}{4\tau_0}.
\label{dustgro}
\ee
Existing models use values of $\tau_0$ ranging between $0.05-30$ Myr \citep{popping2017,vijayan2019,triani2020,dayal2022}. In this work, we adopt a {\it fiducial} value of $\tau_0$ = 1 Myr (see section \ref{model_freeparameters}).

\item The sputtering rate is given by $\dot{M}^{sput}_d = 3 M_d^\prime[\tau_{sput}]^{-1}$ \citep[e.g.][]{McKinnon2017}. Here $\tau_{sput}$ is the dust sputtering timescale that is expressed as \citep{Tsai_1995}
\be
\tau_{sput} = 0.17 \bigg{(}\frac{\alpha_{-1}}{\rho_{-27}}\bigg{)}\bigg{[\bigg(}\frac{T_0}{T_{vir}}\bigg{)^{2.5}+1\bigg{]}} \quad [{\rm Gyr}],
\ee
where $\alpha_{-1}$ is the average grain size of dust in units of 0.1$\mu$m, $\rho_{-27}$ is the ISM density in units of 10$^{-27}$ g cm$^{-3}$, T$_0$ = 2$\times$10$^6$ K the temperature above which the sputtering rate flattens, and $T_{vir}$ is the virial temperature of the host dark matter halo. Assuming a constant dust sputtering rate for our assumed time-step of $t=30$ Myrs, we obtain
\be
M_{d}^{sput} = 45\,\tau^{-1}_{sput}\,M_d{^\prime}.
\label{dustsput}
\ee
\end{itemize}
 
Combining the Eqns. \ref{dustprod}, \ref{dustdes}, \ref{dustej}, \ref{dustgro} and \ref{dustsput} yields a relation between the stellar and dust mass:
 \begin{eqnarray}
 \begin{split}
 M_d  & =  M_* \,f_{rs}\bigg(0.0189 -0.0363 f_* -0.0189 \frac{f_*f_w}{(1-f_*)} \frac{v_s^2}{v_c^2}\bigg)\\
 &\quad \quad \quad 
+ (1-0.0189\,f_{rs}\,f_*)\frac{3M_{d}^\prime}{4\tau_0} -45\tau_{sput}^{-1}M_d^{\prime}.
\end{split}
\label{master}
 \end{eqnarray}

Given that $f_{rs}$ is purely a function of redshift, and $\tau_{sput}$ is a function of redshift and halo mass alone, {\it we have expressed the dust mass-stellar mass relation at a given redshift in terms of only two free parameters - $f_*$ and $\tau_0$.} 

\subsection{The impact of dust on the observed UV luminosity}
We use a single grain size and material density of $a= 0.05\, \mu m$ and $s= 2.25\, {\rm g\, cm^{-3}}$, respectively, assuming graphite/carbonaceous grains \citep{todini2001, nozawa2003}. The dust mass calculated for a given stellar mass above can then be used to compute the ISM optical depth, $\tau_c$, to UV photons as 
\be
\tau_c = \frac{3 M_{d}}{4 \pi r_d^2 a s}.
\ee
We have assumed the extinction cross section of the grains to be $Q_{\rm ext}\approx 1$ at 1500\AA. In principle, for a case where the gas and dark matter halo angular momenta equalise, the gas radius ($r_g$) is linked to the virial radius ($r_{\rm vir}$) as $r_g = 4.5 \lambda r_{\rm vir}$ \citep{ferrara2000} where the spin parameter has a value of $\lambda =0.04$ \citep{davis2009, dayal2018}. However, recent ALMA observations hint at gas radii that are roughly constant for $z \sim 4-7$, for galaxies with stellar masses 8.4 $<$ log$M_*$ $<$ 11.0 (8.6 $<$ log$M_*$ $<$ 10.1) and UV Magnitudes in the range $-23.3 \lesssim \muv \lesssim -19.2$ ($-23 \lesssim \muv \lesssim -21.4$) at $z\sim 4.5 - 6$ ($z \sim 7$), implying that gas fills an increasing fraction of the halo with increasing redshift \citep{fudamoto2022}. We therefore leave the normalisation to be a free parameter such that $r_g = 4.5 \lambda \alpha(z) r_{\rm vir}$. Assuming gas and dust to be co-spatial, the dust radius $r_d = r_g$. 

This optical depth can be easily converted into a value for the escape fraction of continuum photons, $f_c$, by modelling the galaxy as a sphere in which dust and stars are intermixed such that $ f_c = e^{-\tau_c}$. Then, for a total star formation rate value of $\psi$, the SFR in the UV can be calculated as $\psiuv = f_c \psi$ with the infrared star formation rate $\psi_{\rm IR} = (1-f_c) \psi$. We caution that the dust mass and its distribution radius are degenerate in determining UV attenuation. The observed UV luminosity at 1500\AA~in the rest frame then is
\be
L^{obs}_{\rm 1500} (z) = 10^{33.404} f_* \bigg(\frac{\Omega_b}{\Omega_m}\bigg)~2^{- M_{\rm crit} / M_h} f_c(f_*, M_h, z),
\ee
i.e. for a given redshift, the observed UV luminosity depends on the halo mass, the star formation efficiency and the dust attenuation. 

\begin{figure*}
\begin{center}
\center{\includegraphics[scale=0.25]{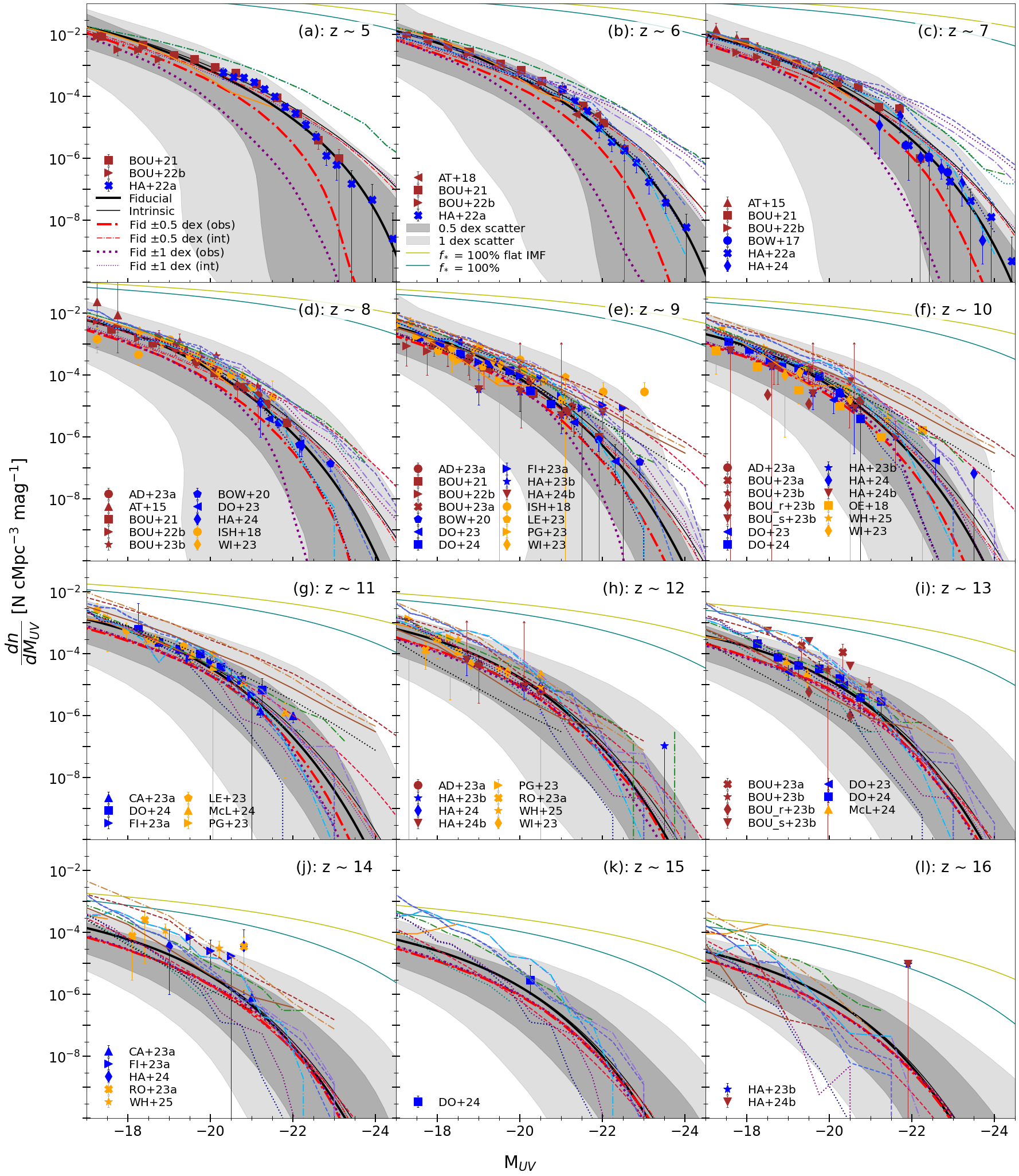}} 
\caption
{The evolving UV LF at $z \sim 5 - 16$. In each panel, lines show theoretical results for intrinsic and dust-attenuated UV LFs for the {\it fiducial} model as well as those allowing for a scatter of $0.5$ and 1 dex on $f_*$, as marked in panel a; the dark and light shaded areas show the corresponding $1 \sigma$ scatter for the $f_*\pm0.5$ and $f_*\pm 1$ dex cases, respectively. In every panel, the solid green and yellow lines show the theoretical upper limits i.e. the ``maximal intrinsic UV LF" for a Salpeter and flat-ish IMF, respectively. Finally, points show observational data, as marked, from \citet{atek2015}, \citet{bowler2017}, \citet{Atek2018}, \citet{ishigaki2018}, \citet{bowler2020}, \citet{bouwens2021}, \citet{Bouwens_2022b}, 
 \citet{bouwens2022jwst}, \citet{harikane2022a}, \citet{donnan2022}, \citet{Willot2023}, \citet{Adams2023a}, \citet{Casey2023a}, \citet{Finkelstein2023a}, \citet{Bouwens2023b}, \citet{Harikane2024b}, \citet{Harikane2023b}, \citet{Harikane2024b}, \citet{Leung2023}, \citet{Robertson2023a}, \citet{PerezGonzalez2023}, \citet{Harikane2024}, \citet{donnan2024jwst} and \citet{Mcleod2024}, \citet{Whitler2025}, as marked in the panels. We also show a number of theoretical predictions: the attenuation-free analytic model from \citealt{ferrara2022} (crimson, dashed). The intrinsic (purple) and observed (blue) UV LF predictions of a synergy of DELPHI and SPHINX \citep{Mauerhofer2025} are shown, each in three variants: an evolving IMF (eIMF, dashed),
an evolving star-formation efficiency (eSFE, dash-dotted), and the fiducial model (dotted). The semi-analytic Santa Cruz models \citep{Somerville2025} are displayed for two choices of the cloud dust-covering fraction, $f_{\rm cloud,d}=0.1$ (sienna, solid) and $f_{\rm cloud,d}=0.5$ (peru, dash-dotted), together with a starburst-enhanced
variant (brown, dashed) and a Kennicutt-Schmidt prescription (\citealt{KSlaw}; black, dotted). We also show results from the ASTRAEUS semi-numerical model from \citep{Hutter2025AstraeusX} with an evolving IMF (green, dash-dotted) and a Salpeter IMF (first presented in \citealt{Cueto2024}; teal, dotted), and the hydrodynamic FIREbox$^{\mathrm{HR}}$ simulation results (\citealt{Feldmann2025}; orange, solid). } 
\label{fig_uvlf}
\end{center}
\end{figure*}
\subsection{Calibrating the free parameters of the model}
\label{model_freeparameters}
The construction of this model rests on a number of underlying assumptions detailed above, for which we have adopted representative or average values from the literature. This allows us to isolate and explore the impact of the three principal free parameters of this model; the star formation efficiency ($f_*$), the timescale associated with grain growth ($\tau_0$) and the gas-to-virial radius relation ($\alpha$). The {\it fiducial} values of these parameters are obtained as follows:
\begin{itemize}
\item $f_*$ is calibrated by fitting to UV LF data at an absolute UV magnitude of $\muv \sim -18$ at $z \sim 5-13$. We use this magnitude since it corresponds to halos that are massive enough to be unaffected by feedback (from either supernovae or reionization) whilst having a low enough mass that dust does not have a reasonable effect on the UV luminosity. We find a redshift evolution of $f_*$ that is best fit by 
\be
f_*(z) = 10^{0.13z-3.5}.
\label{sfeff}
\ee
At this point, the model yields the intrinsic UV LF at all $z \sim 5-20$.

\item Following the work of \cite{dayal2022}, we explore $\tau_0$ values ranging between 0.3 and 30 Myr. While a value of $\tau_0$ = 0.3 Myr (with its extremely effective growth), leads to an overestimation of the observed dust-to-stellar mass ratios compared to observational results from the ALPINE and REBELS surveys, dust growth is completely negligible for $\tau_0 $ = 30 Myr, leading to an under-prediction of this relation. Our model is in good agreement with the observed dust-to-stellar mass ratios (see Sec. \ref{results-dustm-stellm}) for $\tau_0$ values in the range 0.5 - 10 Myr - we assume a fiducial value of $\tau_0 = 1$ Myr. At this point, we obtain the total dust mass for each object.

\item Once the dust mass is obtained, $\alpha(z)$ is determined by matching to the bright end ($\muv \lsim -20$) of the UV LF at $z \sim 5-13$ which yields
\be
\alpha = \bigg{(}\frac{1+z}{11.37}\bigg{)}^{2.46}.
\label{alpha}
\ee
\end{itemize}

\section{Confronting theory and observations in the first billion years}
\label{results}
We now show how our model results compare against the latest data-sets accumulated by JWST and ALMA in the first billion years of the Universe. In addition to the {\it fiducial} model discussed above, we run our model allowing a (gaussian) scatter of 0.5 dex and 1 dex on the star formation efficiency. In these two cases, every halo has a slightly different value of the resulting stellar mass that induces an associated scatter in the dust masses, optical depths and hence the observed UV luminosity, as discussed in what follows.   

\subsection{Redshift evolution of the UV LF and UV luminosity density}
\label{results-uvlf}
We start by discussing the redshift evolution of the UV luminosity function shown in Fig. \ref{fig_uvlf}. As noted in Sec. \ref{model_freeparameters} above, the star formation efficiency ($f_*$) has been calibrated to match to the UV LF  at $\muv \sim -18$ at $z \sim 5-13$. This results in a value of $f_*$ that increases with redshift from about 0.14\% at $z \sim 5$ to $\sim 2.3\%$ by $z \sim 13$. This implies that galaxies of a given {\it intrinsic} luminosity are hosted by lower mass halos with increasing redshift. This calibration also yields the dust mass of every galaxy - we then obtain $\alpha(z)$ (Eqn. \ref{alpha}) that is best required to fit the bright end ($\muv \lsim -20$) of the UV LF at $z \sim 5-13$. This results in a dust radius that increases from about $0.06 r_{\rm vir}$ at $z \sim 5$ to $0.3 r_{\rm vir}$ at $z \sim 13$. Finally while the UV light of galaxies with $\muv \lsim -20$ can have a significant contribution from black hole accretion \citep[e.g.][]{ono2018,piana2022, harikane2022}, in the interest of simplicity, we assume all of the UV light to be purely attributed to star formation in this work. 

Starting with $z \sim 5$ (panel a of Fig. \ref{fig_uvlf}), in the {\it fiducial} model, the intrinsic UV LF increasingly over-predicts the observations at $\muv \lsim -22.5$. Including the effects of dust attenuation brings this model in accord with the data, resulting in UV escape fraction values of $\fescuv \sim 83\% ~(69\%)$ for galaxies with intrinsic magnitudes of $\muv \sim -20 ~(-22)$. We then study the cases where the value of $f_*$ is scattered by $\pm$0.5 and 1 dex at $z \sim 5$, as shown in the same panel. The intrinsic UV LFs in these cases are essentially the same as that in the {\it fiducial} model since the increase/decrease in the star formation efficiency averages out for a given number density. Including the effects of dust attenuation, leads to the following behaviour: for a given halo, a decrease in the star formation efficiency scatters the object into a fainter UV luminosity bin. This is mostly driven by a decrease in the intrinsic UV luminosity although the lower dust mass allows a higher $\fescuv$. An increase in the star formation efficiency leads to a more complicated picture: for halos with $M_h \lsim 10^{10.6}\msun$, an increase in $f_*$ results in objects that are brighter in terms of their observed UV luminosity. This is because they show both an increase in the {\it intrinsic} UV as well as efficient ejection of dust (and gas) in SNII-powered outflows which reduces the dust optical depth. For more massive halos, however, dust ejection is effective, leading to the same effect as above, only if the star formation efficiency is $\gsim 10\,f_*$. For lower $f_*$ values, the observed luminosity of these massive halos actually decreases due to an overall increase in the dust mass and its associated UV attenuation. This leads to an overall {\it decrease} in the average UV LF on the addition of scatter for massive halos as seen from this plot. We note, however, that as a result of the scatter on $f_*$, we find a huge scatter in terms of $\muv$ for a given number density. For example, a number density of $10^{-4}~{\rm cMpc^{-3}}$ encompasses galaxies with $\muv \sim -19$ to $-$21.5 ($-$18 to $ -$22.5) for 0.5 (1) dex of scatter.

\begin{figure}
\begin{center}
\center{\includegraphics[scale=0.4]{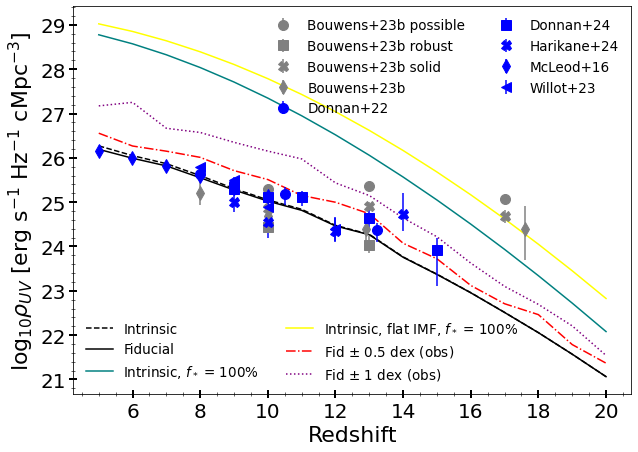}} 
\caption{The redshift evolution of the UV luminosity density ($\rho_{\rm UV}$). As marked, the short dashed and solid lines show the intrinsic and dust-attenuated values of $\rho_{\rm UV}$ in the {\it fiducial} model; the dot-dashed and dotted lines show models allowing for 0.5 and 1 dex of scatter on $f_*$ respectively. The solid green and yellow lines show the theoretical upper limits in the ``maximal" model for a Salpeter and flat-ish IMF, respectively. All of the theoretical models have been integrated down to galaxies with $\muv \lsim -17$ to be able to compare to observations. As marked, points show observational results from \citet{mcleod2016}, \citet{Bouwens2023b}, \citet{donnan2022}, \citet{Willot2023} and \citet{donnan2024jwst}. }
\label{fig_uv_lum_dens}
\end{center}
\end{figure}

While qualitatively the same trends persist out to $z \sim 13$, there is a critical difference in that the impact of dust attenuation decreases with increasing redshift for all galaxies. This is because firstly, for a given stellar mass, the reverse shock destruction term starts playing an increasingly important role with increasing redshift in decreasing the dust mass. Secondly, in our model, both dust and gas occupy a larger fractional volume of the halo with increasing redshift (see Eqn. \ref{alpha}). These effects lead to a severe drop in the dust optical depth to UV photons. At $z \sim 9$ and $13-16$ the observed UV LF inferred by a number of groups \citep{ishigaki2018,bowler2020,harikane2022,bouwens2022jwst,Casey2023a,Harikane2024,Harikane2024b, Harikane2023b} is much more compatible with a model wherein galaxies form stars at 10 times the {\it fiducial} star formation efficiency (as seen from the light shaded area).

We also compare these results to a ``maximal" model where each halo is assumed to have a cosmological baryon-to-dark matter ratio and all of this gas is allowed to form stars instantaneously (i.e. $f_*=100\%$). We then calculate the UV luminosity for two different IMFs: {\it (i)} the first uses our standard ($1-100\msun$) Salpeter IMF wherein $L^{int}_{1500} = 10^{33.404} (M_*/\msun) ~ {\rm erg\, s}^{-1} {\rm \AA}^{-1}$ ; {\it (ii)} the second uses a flat-ish IMF with a slope of 0.1 yielding a 5.5 times higher value of $L^{int}_{1500} = 10^{34.148} (M_*/\msun) ~ {\rm erg\, s}^{-1} {\rm \AA}^{-1}$. Reassuringly, as of now, at all $z \sim 5-15$, these upper limits are appreciably above the observed UV LF implying that no observations of the UV LF pose any concern to the standard ($\Lambda$CDM) cosmological model. If confirmed, the bright end of the UV LF at $z \sim 16$ poses a challenge in that it lies between the Salpeter and flat-ish IMFs even in the ``maximal" model.

We then discuss the theoretically-inferred UV luminosity density as shown in Fig. \ref{fig_uv_lum_dens}. In the {\it fiducial} case, integrating down to $\muv = -17$, the UV luminosity density declines by 5 orders of magnitude from  ${\rm log}(\rhouv) \simeq 26.2$ at $z \sim 5$, to ${\rm log}(\rhouv) \simeq 21$ by $z \sim 20$. At all redshifts, $\rhouv$ is dominated by low-mass systems ($M_* \sim 10^{6.2-7}~\msun$) with intermediate magnitudes ($\muv \sim$ -17 to -19). These are hosted in halos of $M_h \sim 10^{10.1-10.8}~\msun$ at $z \sim 5$, decreasing to $M_h \sim 10^{9.3-9.5}~\msun$ by $z \sim 20$. Given their low mass (and associated dust contents), these systems are essentially unaffected by dust attenuation.   

\begin{figure*}
\center{\includegraphics[scale=0.45]{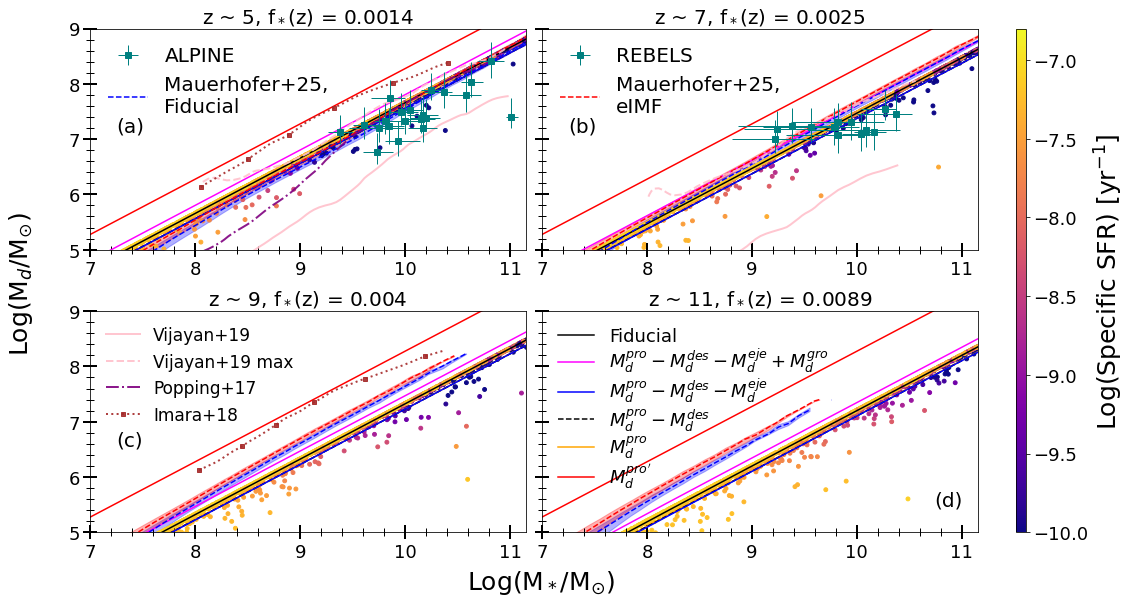}} 
\caption{The dust mass as a function of the stellar mass at $z \sim 5-11$, as marked, colour-coded by the observed specific SFR, $\psiuv/M_*$. The solid (black) line shows the results of the {\it fiducial} model. The other lines show this relation successively accounting for the processes of dust production, reverse shocks, dust destruction, dust ejection and grain growth in the ISM. Filled circles are obtained by scattering the star formation efficiency ($f_*$) by $\pm$ 1 dex. Filled squares show observational data from the ALPINE survey \citep{fudamoto2020} at $z \sim 5$ and from the REBELS survey \citep{bouwens2022} at $z \sim 7$, in panels a and b, respectively. The blue dashed lines and shaded areas represent the fiducial model in \citet{Mauerhofer2025} at the respective $z$, while orange dashed lines and shaded areas correspond to the same model, adopting an IMF evolving with $z$. We also include results from other simulations: dash-dotted purple lines are trends computed by \citet{popping2017}, pink solid and dashed lines are 
the fiducial and maximal models in \citet{vijayan2019} respectively, while brown square-dotted lines correspond to the results of \citet{Imara2018}.}
\label{fig_dust_m_stell_m_cmap}
\end{figure*}

As noted above, our {\it fiducial} model has been calibrated to match to the observed UV LF at $z \sim 5-13$ which also results in matching to the observed UV luminosity density (integrated from the UV LF) at these redshifts. We note that a 0.5 (1) dex scatter on $f_*$ leads to a value that is about 0.4 (1.2) dex higher than the {\it fiducial} model at $z \sim 5-12$ driven by objects up-scattered in terms of $f_*$. This results in these models over-predicting the observations, especially at $z \lsim 10$. However, given the dispersion on the observational results at $z \gsim 13$, all of these models are compatible with the data at early epochs. Interestingly, the tentative $\rhouv$ estimates at $z \sim 18$ \citep{Bouwens2023b} are more than 1 dex higher than the values predicted by any of these models and even exceed the ``maximal" model for a Salpeter IMF. In fact, these points are more compatible with the ``maximal" model that assumes a flat-ish IMF. However, the nature of such ultra-high redshift sources remains debated as of now \citep[e.g.][]{popping2022}.

Finally, we compare our results to other works including: (i) the analytic attenuation-free model \citep{ferrara2022}; (ii) semi-analytic models including the Santa-Cruz \citep{Somerville2025} and {\sc delphi} models \citep{Mauerhofer2025}; (iii) semi-numerical models such as {\sc astraeus} \citep{Hutter2025AstraeusX}; and the {\sc firebox} hydrodynamic simulations \citep{Feldmann2025}. As seen, these models are very compatible with our results within error bars. Crucially, we note that the dispersions between these results increase with redshift: while most model results are in reasonable agreement with each other at $z \lsim 8$, they show an increasingly different behaviour with increasing redshifts, driven by the impact of their different volumes simulated, varying dust prescriptions, star formation efficiencies and IMFs used. Our results, including the impact of (1 dex) scatter on the star formation efficiency are in reasonable agreement with most other theoretical works at $z \gsim 11$, for $\muv \lsim -19$; at lower luminosities, the theoretical models we compare to show a large dispersion, with some predicting up to a factor 100 more galaxies at the faint end. Ongoing observations with the JWST will be crucial on shedding light on the faint end of the UV LF and informing models.

\begin{figure*}
\center{\includegraphics[scale=0.45]{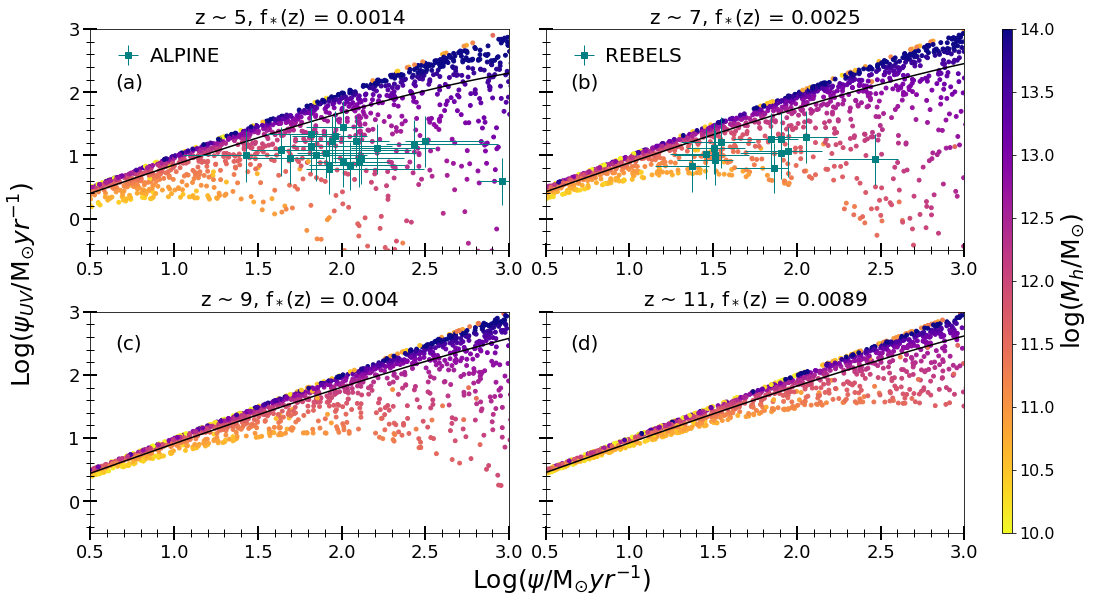}} 
\caption{The SFR inferred from the UV as a function of the total intrinsic SFR for $z \sim 5-11$, as marked, colour-coded by the host halo mass. The solid (black) line shows results from the {\it fiducial} model; filled circles show results obtained by scattering the star formation efficiency ($f_*$) by $\pm$ 1 dex. Square points show observational results from the ALPINE survey \citep{fudamoto2020} at $z \sim 5$ and from the REBELS survey \citep{bouwens2022} at $z \sim 7$. } 
\label{fig_sfr_uv_sfr_tot_cmap}
\end{figure*}

\begin{figure}
\center{\includegraphics[scale=0.4]{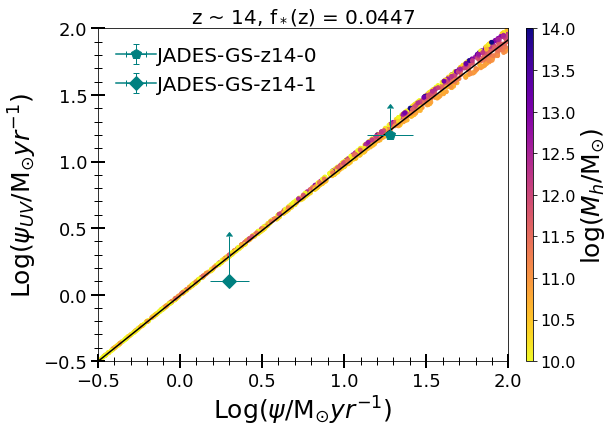}} 
\caption{The SFR inferred from the UV as a function of the total intrinsic SFR for $z \sim 11$, colour-coded by the host halo mass. In each panel, the solid (black) line shows results from the {\it fiducial} model; filled circles show results obtained by scattering the star formation efficiency ($f_*$) by $\pm$ 1 dex. The teal data points correspond to JADES-GS-z14-0 and JADES-GS-z14-1  \citep{Carniani2024}, as shown in the legend.} 
\label{fig_sfr_uv_sfr_tot_z14}
\end{figure}
\subsection{Relation between the dust and stellar mass}
\label{results-dustm-stellm}
We now discuss the key dust processes that drive the relation between the dust and stellar masses at $z \sim 5-11$, as shown in Fig. \ref{fig_dust_m_stell_m_cmap}. At $z \sim 5$ (panel a), in the {\it fiducial} model the dust and stellar masses are related as ${\rm log}(M_d) = {\rm log}(M_*) - 2.35$ i.e. a galaxy with $M_* \sim 10^{9.5}\msun$ has a total dust mass of $M_d \sim 10^{7.15}\msun$. From the same plot, we see that dust production alone ($M_d^{\rm pro'}$) yields a dust mass of $\nu M_* =0.0189 M_* [\msun]$ i.e. a stellar mass of $M_* \sim 10^{9.5}\msun$ can produce a dust mass $\sim 10^{7.7}\msun$. However, only about $19\%$ of this dust survives when including the impact of reverse shock destruction resulting in $M_d^{\rm pro}=10^7 \msun$. The process of dust destruction ($M_d^{\rm des}$) is largely ineffective at this redshift with  $M_d^{\rm des} \simeq 10^{-4}\,M_d^{\rm pro}$. Galaxies with $M_*\gsim 10^{6}\msun$ correspond to $M_h \sim 10^{10}\msun$ - such deep potentials render ejection essentially ineffective, as seen from the same plot. Sputtering plays a crucial role, decreasing the dust mass by a factor $\sim$1.4. Finally, ISM grain growth more than compensates for this, increasing the dust mass by a factor of $\sim$3.2, resulting in the final linear relation seen.  

We also colour-code the scatter points by their observed specific SFR (sSFR) in the UV i.e., $\fescuv~\psi/M_* = \psiuv/M_*$. Since $\psi \propto L_{1500}$ and $L_{1500} \propto M_*$, the ratio $\psi/M_*$ is a constant in the {\it fiducial} model with the sSFR effectively being a proxy for the UV escape fraction $\fescuv$. In this model, galaxies basically track the average relation for dust attenuation. Adding scatter on $f_*$ changes the situation in that galaxies with a star formation efficiency $\gsim 10\,f_*$ eject and destroy dust more efficiently. These rare, highly star forming objects are scattered to lower dust masses by up to a factor up to 10 below the {\it fiducial} line. We also note the presence of systems below the {\it fiducial} line with observed specific star formation rates (sSFR) $\gsim 10^{-8}$ for stellar masses $M_* \lesssim 10^{9} ~\msun$ at $z \sim 5$. These systems undergo extremely efficient star formation in low mass halos which eject their entire gas and dust contents, resulting in a high $f_{esc}^{UV}$ value. We then compare our model results with the dust-to-stellar mass relations inferred by the ALPINE survey \citep{fudamoto2020} at the same redshift. As seen, the bulk of ALPINE points are in agreement with our {\it fiducial} model, by construction. Observed outliers of $M_* \sim 10^{9.7-10.2}\msun$ with low dust masses ($M_d \sim 10^{6.6-7}\msun$) can be explained by the 1 dex scatter case. This is driven by the fact that these systems have very low values of $\fescuv$ despite their small dust reservoirs with respect to the average population, because dust is distributed more compactly in their host halos. We note that there is an extremely massive outlier in the ALPINE sample ($M_* \sim 10^{11}\msun$) with $M_d \sim 10^{7.4}\msun$ i.e. a dust to stellar mass ratio of 0.02\% which lies outside even the 1 dex scatter case considered here.

Our qualitative results essentially remain unchanged as a function of redshift. The most crucial difference is that the importance of reverse shock increases due to the increasing ISM densities with increasing redshift. As a result, at $z \sim 7$, we find a slightly lower (by 0.2 dex) normalisation of the relation such that ${\rm log}(M_d) = {\rm log}(M_*) - 2.54$. Comparing to observations from the REBELS survey \citep{bouwens2022}, we again find our {\it fiducial} model to be in reasonable agreement with the bulk of the data points. However, the dust-stellar mass relation observed by the REBELS survey is much flatter than that inferred by the ALPINE survey. One possible reason for this could be an under-estimation of the stellar masses for the lowest mass objects due to the assumed constant star formation history as pointed out by \citet{topping_rebels}. On the other hand, if these masses are correct, solutions include e.g. higher dust yields per SN, higher post-reverse shock survival fractions or lower dust growth timescales, to name a few. Allowing for a 1 dex scatter on $f_*$, again, we are able to explain the high stellar mass objects that have low dust masses as dusty and compact sources with a high dust optical depth.

The normalisation of this relation keeps decreasing with redshift such that  ${\rm log}(M_d) = {\rm log}(M_*) - \gamma$ where $\gamma = 2.68~ (2.84)$ at $z \sim 9~(11)$ as shown in panels (c) and (d) of the same figure. Again, allowing for a 1 dex scatter on $f_*$, we find dust masses that are up to a factor 100 below the average relation. 

Finally, we compare our results to those obtained by a number of other theoretical works as shown in the same figure. We start by comparing to the semi-analytic results of \citet{Mauerhofer2025} which are very similar in spirit to the dust model used here - these authors have implemented ISM physical results (in terms of cold gas fractions and star formation efficiencies) from the \code{sphinx-20} hydrodynamical simulations \citep{Rosdahl2022} into the \texttt{DELPHI} semi-analytic model \citep{dayal2014a, dayal2022} to match to the latest results from both ALMA and JWST at $z \gsim 5$. As seen from Fig. \ref{fig_dust_m_stell_m_cmap}, at $z \sim 5-7$ our results are in excellent agreement with those from \citep{Mauerhofer2025}. However, due to a redshift-dependent increase in both the cold gas fraction (that determines the rate of dust growth in the ISM) and star formation efficiency (that determines the dust mass production), required to explain the UV LFs from the JWST, the results from \citet{Mauerhofer2025} over-predict the dust mass at all stellar masses, compared to our models at $z \gsim 9$. However, we note that at a given stellar mass, the dust mass is, at most, higher only by a factor of 3 at $z \sim 11$. We also show comparison with the semi-analytic results of \citet{popping2017} and \citet{vijayan2019}, and the analytic results of \citet{Imara2018}. As seen, these models from \citet{vijayan2019} and \citet{popping2017} are in good agreement with our fiducial results. The results from \citet{Imara2018} (available only at $z \sim 5,9$) are closer to the upper limit (production only) from our model; we caution this analytic model does not explicitly model the dust masses. 

\subsection{Dust impact on observed UV star formation rates}
\label{results-dust impact}
We now examine the effect of dust on the observed star formation rate  of galaxies through the fraction of the total SFR ($\psi$) that would be observed in the UV ($\psiuv = \fescuv \, \psi$). In the {\it fiducial} model, $\psi \propto M_*$ i.e. galaxies of increasing SFR inhabit increasingly massive halos at any redshift. Given that the dust mass increases with $M_*$, it leads to a corresponding decrease in $\fescuv$. This behaviour can be seen from (panel a) of Fig. \ref{fig_sfr_uv_sfr_tot_cmap} where the {\it fiducial} relation predicts $\psiuv \sim 10 ~ (40) ~\msun \,yr^{-1}$ for $\psi \sim 15 ~(80) ~ \msun \, yr^{-1}$, implying $\fescuv \sim 68\% ~(50\%)$. Interestingly, the {\it fiducial} model over-predicts $\psiuv$ for a given SFR as compared to ALPINE results. Indeed, matching to these results requires these objects to form stars with an efficiency that is 5 to 15 times higher than the average value. As seen from the same figure, galaxies can end up in very different parts of the $\psi-\psiuv$ space depending on their $f_*$ value. For example, a down-scattered value of $f_*$ leads to both a lower stellar and dust mass for a given halo as compared to the {\it fiducial} model since $\psi \propto M_* \propto M_d$. This reduced dust reservoir leads to a lower value of the UV optical depth i.e. a higher value of $\psiuv$. On the other hand, increasing the star formation rate of a galaxy increases both its stellar mass and dust content whilst leaving the dust radius unchanged, resulting in a lower value of $\fescuv$ i.e. a lower value of $\psiuv$ - for example, as seen from this plot, we find values of $\fescuv \sim 24-84\%$ for $\psi = 10~\msun\,{\rm yr^{-1}}$ dropping to $\fescuv \sim 0.07 - 19\%$ for $\psi = 10^{3.5}~\msun\,{\rm yr^{-1}}$. Efficiently star forming galaxies (with $f_*$ upscattered by a factor of 10) in low mass halos ($M_h \lesssim 10^{10.6}$) are an exception to the above. In this regime, ejection can push out a high percentage of the dust content, bringing them closer to the $\psiuv \sim \psi$ limit.

\begin{figure*}
\center{\includegraphics[scale=0.5]{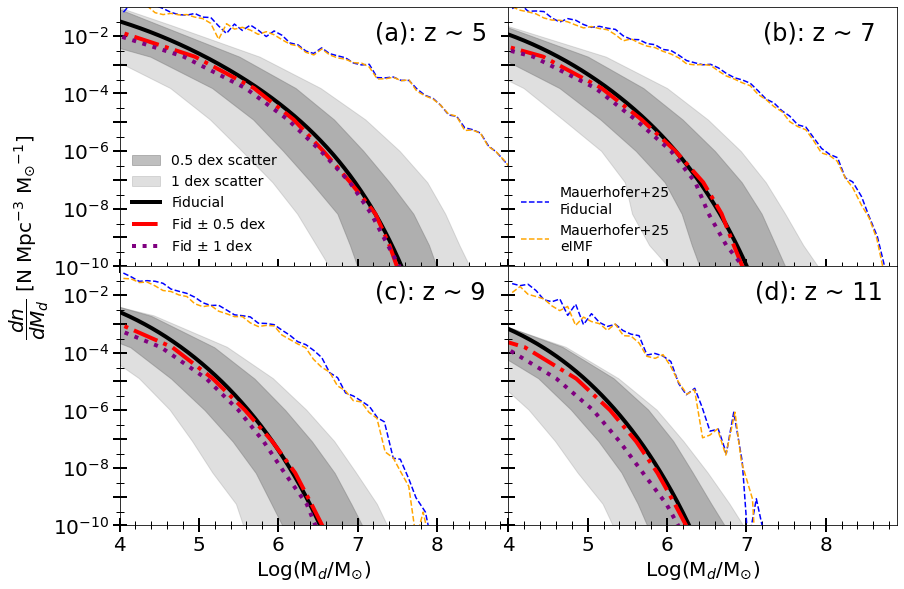}} 
\caption{The dust mass function for $z \sim 5-11$, as marked. The black line corresponds to the results of the {\it fiducial} model. The red dot-dashed and purple dotted lines show results obtained by scattering the star formation efficiency for every halo within 0.5 and 1 dex, respectively; the light and dark grey shaded areas show the scatter associated with 0.5 and 1 dex of scatter on $f_*$, respectively. The blue dashed lines represent the fiducial model in \citet{Mauerhofer2025} at the respective $z$, while the orange dashed lines correspond to the same model, adopting an IMF evolving with $z$.}
\label{fig_dustmf}
\end{figure*}

In our model, a variation in $\fescuv$ can either be attributed to a change in the dust mass or a more compact/dispersed dust distribution. In the {\it fiducial} model, $\psi \propto M_*$ i.e. galaxies with the same intrinsic SFR, contain approximately the same dust mass. However, on adding 1 dex of scatter on $f_*$, we find a range of $\psiuv$ values for a given $\psi$. As noted above, low-mass halos with up-scattered $f_*$ values show lower $\psiuv$ values than the average relation due to a more compact dust distribution. On the other hand, objects down-scattered in terms of $f_*$ show higher $\psiuv$ values for a given intrinsic SFR. Adding 1 dex of scatter on $f_*$ allows us to explain all of the ALPINE data points at $z \sim 5$ in the $\psi-\psiuv$ plane - we see these sources are hosted in halos of masses $\sim 10^{10.5-12.5}\msun$.

At $z \sim 7$, we find $\psiuv \sim 10~(40)$ for $\psi \sim 13.4 ~(66.5) ~ \msun \, yr^{-1}$ yielding $\fescuv \sim 75\% ~(60\%)$. As noted earlier, galaxies of a given stellar mass (and hence star formation rate) are hosted by lower-mass halos with increasing redshift due to an increase in $f_*$; $f_*=0.25\%$ at $z \sim 7$ compared to $f_* = 0.14\%$ at $z \sim 5$. However, for a given stellar mass, the dust mass decreases with redshift due to destruction in reverse shocks which boosts up $\fescuv$. We again find that the {\it fiducial} model provides an upper limit to the $\psi_{\text{UV}} - \psi$ relation observed by the REBELS survey \citep{bouwens2022} data at $z \sim 7$. Matching to these observations again requires about a dex of scatter on $f_*$ i.e. these are intrinsically bright and highly star forming ($\psi \sim$ 24 $-$ 294 $\msun$ yr$^{-1}$) but highly obscured objects with $\fescuv \sim 3\% - 46\%$. The scatter of the model indicates these galaxies are in dark matter halos with masses similar to ALPINE \citep{fudamoto2020} galaxies at $z \sim 5$ with $M_h \approx 10^{10.8} - 10^{12.4}$ M$_\odot$ and are forming stars with 4 to 10 times higher efficiency than the {\it fiducial} value of $f_*(z = 7) = 0.0025$. We emphasise that these objects are in halos massive enough to keep most of their dust content within their gravitational potential well, despite their increased star formation efficiencies.

We then use our model to predict the $\psiuv-\psi$ relations at $z \sim 9$, $11$ and $14$ as shown in (panels c and d of) the same figure, and Fig. \ref{fig_sfr_uv_sfr_tot_z14} respectively. 
As expected, the {\it fiducial} relation shows a slight increase in $\psiuv$ by about 0.02 dex i.e. $\psiuv \sim 64-67$ for $\psi \sim 100 ~ \msun \, yr^{-1}$ yielding $\fescuv \sim 64\%~(67\%)$ at $z \sim 9~(11)$. With increasing redshift, the dust radius fills an increasing fraction of the virial radius in our model. As a result, $\fescuv$ increases for all $\psi$ even accounting for scatter on $f_*$. For example, by $z \sim 11$, we find a minimum $\fescuv \sim 5.5\%$ for the most highly star forming systems with $\psi \sim 10^{3.5} ~ \msun \, yr^{-1}$ while this value drops to $\fescuv \sim 0.2\%$ for similar sources at $z \sim 9$. 
At $z \sim 14$, the {\it fiducial} model predicts $\fescuv \sim 82\%$ for $\psi \sim 100 ~ \msun \, yr^{-1}$, facilitated by the decreasing effect of dust attenuation with $z$. As shown in Fig.  \ref{fig_sfr_uv_sfr_tot_z14}, including the full range of scatter around the $f_*(z) \sim 0.0447$, only brings galaxies $\sim 0.2$ dex below the $\psiuv \sim \psi$ relation for $\psi \sim 100 ~ \msun \, yr^{-1}$, corresponding to an average $\fescuv = 94.1^{+5.9}_{-6.5}\,\%$ for the illustrated range of SFRs.
We finally test the predictability of our model, by including observations of JADES-GS-z14-0 and JADES-GS-z14-1 \citep{Carniani2024} at $z \sim 14$ in Fig.  \ref{fig_sfr_uv_sfr_tot_z14}. We find that the {\it fiducial} model is consistent with observations of the SFR at $z\sim 14$. We treat the $\fescuv$ measurement in these data points as a lower limit, as \citet{Carniani2024} measure the escape fraction of Lyman Continuum photons ($f_{esc}^{LyC}$) through SED fitting; $f_{esc}^{LyC}$ is affected by both dust attenuation as well as absorption by neutral hydrogen HI, whereas $\fescuv$ at 1500\AA$\,$ is only affected by dust attenuation. In that sense, we always have $f_{esc}^{LyC} < \fescuv$. \citet{Carniani2024} report $f_{esc}^{LyC} = 0.84^{+0.09}_{-0.16}$ and $f_{esc}^{LyC} = 0.63^{+0.25}_{-0.29}$ for GS-z14-0 and GS-z14-1 respectively, with very low dust extinction in both galaxies, consistent with our prediction of dust not affecting the observable properties of galaxies  at $z\sim 14$. The above imply $\fescuv$ very close to 100\% for both galaxies. This is in agreement with our model, as JADES-GS-z14-0 and JADES-GS-z14-1 are consistent with our {\it fiducial} $\psiuv \sim \psi$ relation within 1$\sigma$. The {\it fiducial} model then nicely predicts the SFR of individual galaxies at high $z$. Nevertheless, such indirect measurements should be used with caution, as they are plagued by numerous assumptions related to SED modelling, e.g. the adopted star formation history, initial mass function, dust geometry etc.

\subsection{The dust mass function and its redshift evolution}
\label{results-dustmassfn}

\begin{figure*}
\center{\includegraphics[scale=0.4]{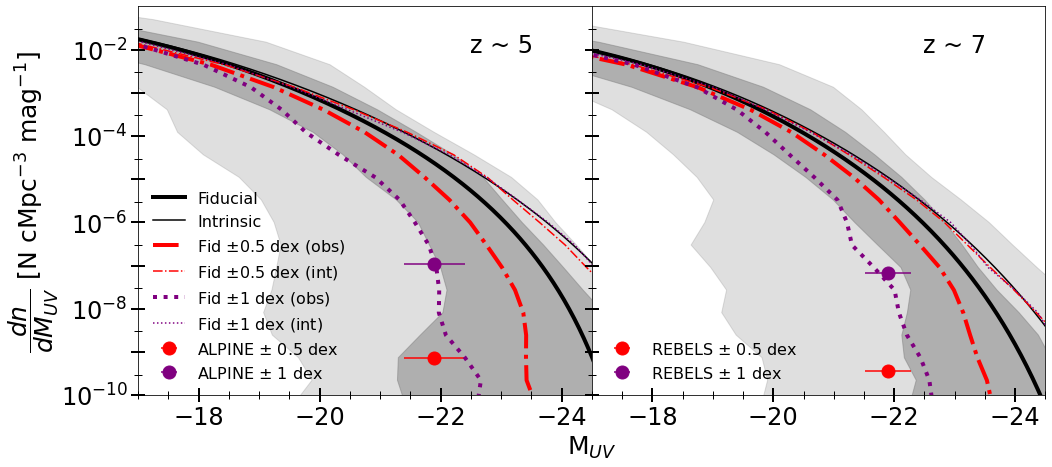}} 
\caption{Plot of the UV LF at $z \sim 5$ (left) and $z \sim 7$ (right), including the UV LF data points inferred from the ALPINE and REBELS surveys allowing for 0.5 and 1 dex of scatter on $f_*$. As marked, we show intrinsic and dust-attenuated results from the theoretical model for the {\it fiducial} model as well as models where $f_*$ is scattered by 0.5 and 1 dex; the dark and light grey shaded regions show the $1\sigma$ scatter associated with the last two cases.} 
\label{fig_uvlf_wALMA}
\end{figure*}

We now explore the dust mass function, i.e. the number density of galaxies as a function of their dust mass, at $z \sim 5-11$ as shown in Fig. \ref{fig_dustmf}. In the {\it fiducial} model, the dust mass tracks the halo mass as ${\rm log}(M_d) = {\rm log}(0.146\,f_*(z)\,2^{-M_c/M_h} M_h) - \gamma$ where $\gamma$ is the ratio between the dust and stellar mass discussed in Sec. \ref{results-dustm-stellm}. For a halo mass of $M_h = 10^{10}\msun$ corresponding to a number density of  $\sim 10^{-1.2}~(10^{-3.1})~{\rm cMpc^{-3}}$ at $z \sim 5~(11)$, this results in $M_d = 10^{3.66} ~(10^{3.97})\msun$. As expected, the increase in $f_*$ with increasing redshift leads to an overall increase in the dust mass for a given halo despite the higher destruction in reverse shocks. However, this increase is compensated by the decrease in both the amplitude and knee-mass of the halo mass function with increasing redshift which is reflected in overall lower amplitude of the dust mass function. We can also consider the dust mass function at a given number density (corresponding to decreasing halo mass with increasing redshift) - in this case, for a number density of $10^{-4}~{\rm cMpc^{-3}}$, the dust mass decreases from $10^{5.8}$ M$_{\odot}$ at $z \sim 5$ to $10^{4.6}$ M$_\odot$ by $z \sim 11$. 

As discussed in the previous subsections, increasing (decreasing) the star formation efficiency of a halo at a given redshift produces more (less) dusty systems, except for cases of significantly increased $f_*$ compared to the average population, e.g. $\gsim 10~f_*(z)$ at $z \sim 5$. Given the rarity of the above cases, the effect of varying $f_*$ averages out (similar to the intrinsic UV LF), resulting in average dust mass functions that match quite well in the {\it fiducial} and $f_*\pm$ 1 dex cases at $z \sim 5-7$. Nevertheless, the 1 dex scatter case predicts 0.3 dex (0.5 dex) lower dust masses as compared to the {\it fiducial} model at $z \sim 9 ~(11)$ mostly driven by a heightened effectiveness of dust destruction.

We now compare our dust mass function to the results of \citet{Mauerhofer2025}. Our work predicts up to $\sim2$ dex lower dust mass for halos of the same number density at $z \sim 5$, with the magnitude of the difference lowering to $\sim1$ dex at $z \sim 11$. However, both models reproduce the observed dust-to-stellar mass relations at $z \sim 5$ and 7, although they suggest a different evolution of this relation across cosmic time, as seen in section \ref{results-dustm-stellm}. Since the halo mass functions used in both models are the same, and the dust-to-stellar mass relations of the two models match, the above discrepancy arises from the varying halo mass-stellar mass relations in these two models, i.e. the different amplitudes of the stellar mass functions. We discuss this further in Appendix A.
\subsection{Obscured galaxies at high redshift: outliers or representative of the average population?}
\label{discussion-datapoints}

We now contextualise the objects being observed by the ALPINE and REBELS surveys at $z \sim 5-7$. As seen from the discussions above, while the dust masses and stellar masses observed are well fit by the {\it fiducial} model, it over-predicts the escape fraction of UV photons compared to these observations. In order for this tension to be reconciled, our model predicts that these galaxies are forming stars with a star formation efficiency higher by 5-15 times the {\it fiducial} value and inhabit halos with $M_h \sim 10^{11.3 - 12.9}~(10^{10.8 - 12.4})~\msun$ at $z \sim 5$ ($z \sim 7$). This implies these  galaxies are not representative of the average population of galaxies that make up the UV LF \citep[see also discussion in][]{dayal2022}. 

To better illustrate the above argument, we generate UV LF data points for ALPINE and REBELS observations, using their estimates of the observed UV SFR and stellar mass compiled in previous works \citep{fudamoto2020, dayal2022}. We start by using the UV SFR to obtain their $\muv$ values. We then use three cases for the star formation efficiency ($f_*^{'}$) of these objects: {\it (i)} Case A: the {\it fiducial} $f_*^{'}= f_*(z)$ as given in Eqn. \ref{sfeff}; {\it (ii)} Case B: $f_*^{'} = 10^{0.5}~f_*(z)$ and {\it (iii)} Case C: $f_*^{'} = 10^{1.0}~f_*(z)$. We combine this $f_*^{'}$ value with the observed stellar mass to calculate the gas mass as $M_g = M_*/f_*^{'}$. This is used to infer a related halo mass (using Eqn. \ref{mg}) and the associated number density from the halo mass function. The resulting UV LF data points are then generated by calculating the average $\muv$ and the standard deviation of objects for a given number density bin. 

We plot these estimates of the UV LF against our model results in Fig. \ref{fig_uvlf_wALMA}. We find that the ALPINE and REBELS data points would, on average, inhabit extremely massive halos with $M_h \sim 10^{14}\msun$ and $M_h \sim 10^{13.4}~\msun$ respectively (number densities $\sim 10^{-12.6}~{\rm cMpc^{-3}}$ and $\sim 10^{-12.8}~{\rm cMpc^{-3}}$) for Case A above; for this reason these data points are not shown in Fig. \ref{fig_uvlf_wALMA}. Adopting increasingly high star formation efficiencies would allow such galaxies to be hosted in less massive halos. Indeed, we find that the ALPINE points can be accommodated in halos of $M_h \sim 10^{13.4}$ and $10^{13}\msun$ (with number densities $\sim 10^{-9.2} ~{\rm and}~10^{-7}~{\rm cMpc^{-3}}$) allowing for a 0.5 dex (case B) and 1 dex (case C) of scatter on $f_*$, respectively. The REBELS points would correspond to similar mass halos of $M_h \sim 10^{12.9}~(10^{12.4})~\msun$ with number densities $\sim 10^{-9.5}~(10^{-7.2})~{\rm cMpc^{-3}}$ allowing 0.5 (1) dex of scatter on $f_*^{'}$. Finally, in Figure \ref{fig_uvlf_wALMA}, only Case C generated
data points match their respective UV LF indicating these galaxies lie significantly above the main sequence of star formation. Thus, these objects are not representative of the average population of galaxies that make up the UV LF at these early epochs.

\section{Conclusions and Discussion}
\label{conclusions-discussion}
In this work, we develop an analytic formalism to link the stellar masses (and star formation rates) of early systems, at $z \gsim 5$, to their dust enrichment and the associated UV attenuation. We account for the key dust processes of dust production in SNII, destruction and ejection by SN shocks, thermal sputtering in the warm ISM and grain growth in the cold ISM. The dust-to-stellar mass ratios from these calculations are baselined against ALMA observations at $z \sim 5$ and $7$ from the ALPINE and REBELS surveys, respectively. We explore star formation rates that can vary by an order of magnitude around the ``main" sequence (linking the SFR and stellar mass) and their impact on the dust enrichment and attenuation as a means of explaining the over-abundant, seemingly dust-free bright systems revealed by JWST at $z \gsim 11$. The key results of this work are: 

\begin{itemize}
    \item Matching to the evolving UV LF requires a star formation efficiency that evolves as $f_*(z) = 10^{0.13z-3.5}$ and a dust distribution radius that increases with redshift from $0.06 r_{\rm vir}$ at $z \sim 5$ to $0.3 r_{\rm vir}$ by $z \sim 13$. A combination of efficient destruction in reverse shocks and an increasing dust radius result in the impact of dust decreasing with increasing redshift for all halo masses. We also compare our results to a number of other theoretical works, which, despite showing a reasonable agreement for $\muv \lsim -19$ systems, show a UV LF that can vary by more than 2 orders of magnitude at the faint end. 
    
\item A scatter on $f_*$ leads to a mass-dependent impact on the UV LF: while for a given halo, a decrease in the star formation efficiency scatters the object into a fainter UV luminosity bin, an increase in $f_*$ leads to a more complicated picture. While low-mass objects ($M_h \lsim 10^{10.6}\msun$ at $z \sim 5$) up-scattered in terms
of star formation are brighter (due to an increase in the intrinsic UV luminosity and efficient dust ejection), higher mass objects appear fainter since they can maintain their dust mass unless the star formation efficiency$\gsim 10f_*(z)$.

\item At $z \sim 9-16$ the observationally-inferred UV LF is much more compatible with a model wherein galaxies form stars at 10 times the {\it fiducial} star formation efficiency. 

\item The UV luminosity density increases with cosmic time as more and more systems assemble and undergo star formation. At all redshifts, $\rhouv$ is dominated by low-mass systems ($M_* \sim 10^{6.2-7}~\msun$) with intermediate magnitudes ($\muv \sim$ -17 to -19). While our results are in agreement with the observationally inferred UV luminosity density at $z \sim 5 - 13$, the estimates at $z \sim 18$ \citep{Bouwens2023b} pose a challenge, exceeding even our ``maximal" model ($f_*=100\%$) for a Salpeter IMF. In fact, these points are more compatible with the ``maximal" model that assumes a flat-ish IMF suggesting a need for caution in interpreting data at such high redshifts.

\item The {\it fiducial} model yields a dust-to-stellar mass relation given by log$M_d$ = log$M_*$ - $\gamma$, with $\gamma = -2.35$ at $z \sim 5$. The slope of the dust-to-stellar mass relation remains unchanged with redshift, with the normalisation decreasing with redshift (to -2.84 by $z \sim 11$) mainly due to increased destruction by reverse shocks. The dust enrichment of early systems is driven by dust production in SNII ejecta. Grain growth and sputtering are the second and third most crucial processes; ejection and destruction are inefficient except in cases of highly increased $f_*$. Our results are also in reasonable agreement with those from e.g. \citet{popping2017} and \citet{vijayan2019}.
    
\item Our model aligns well with ALPINE and REBELS survey observations at $z \sim 5$ and $7$ when invoking efficient dust growth ($\tau_0$ = 1 Myr). Outliers of the relation at $M_* \sim 10^{9.7 - 10.2}~\msun$ with $M_d \sim 10^{6.6-7}~\msun$ can be explained by star formation efficiencies scattered around the average relation by 1 dex.

\item The {\it fiducial} model predicts that at $z \sim 5$ ($z \sim 7$), for a $\psiuv$ of 10 and 40 $M_{\odot}\, \text{yr}^{-1}$, the intrinsic SFR is 15 and 80 $M_{\odot} \, \text{yr}^{-1}$ (13.4 and 66.5 $M_{\odot} \, \text{yr}^{-1}$), corresponding to $\fescuv$ values of 68\% and 50\% (75\% and 60\%). The {\it fiducial} $\psiuv - \psi$ relation shows a 0.02 dex increase in $\psiuv$ for the same $\psi$ from $z \sim 9$ to $z \sim 11$. The {\it fiducial} $\psiuv - \psi$ relation matches observations at $z \sim 14$.

    \item The {\it fiducial} model over-predicts $\psiuv$ for a given SFR in the context of both the ALPINE and REBELS surveys, requiring a factor of 5-15 increase in $f_*$ to match the results. These galaxies are predicted to reside in halos of mass $M_h \approx 10^{10.8 - 12.4}~\msun$. 
 
    \item The overall amplitude of the dust mass function decreases with increasing redshift, driven by a combination of dust destruction in reverse shocks and the evolution of the halo mass function. For e.g., halos with a number density of $10^{-4}~{\rm cMpc^{-3}}$ host $M_d \sim 10^{5.8}~\msun$ at $z \sim 5$, dropping to $M_d \sim 10^{4.6}~\msun$ by $z \sim 11$.The dust mass function in this work should be treated as a lower limit.
    
    \item ALMA observations of UV bright galaxies at $z \sim 5$ and 7 are not representative of the average population of galaxies that make up the UV LF; these galaxies seem to be forming stars with an efficiency that is about 10 times greater than the average $f_*$ value at that redshift. 
\end{itemize}

Overall, our results allow early galaxies to be dust rich. However, dust plays an increasingly limited role in regulating the UV luminosity since it covers a larger volume fraction of the host halo with increasing redshift. We note this is contrary to other theoretical works \citep{ferrara2022,2023MNRAS.520.2445Z} that find early galaxies to essentially be dust-free systems. Observables predicted by our model, such as the dust mass function, will be crucial in shedding light on this issue with forthcoming 
ALMA observations.  

We end this work by discussing some key caveats in our formalism. Firstly, we have assumed gas, metals, and dust to be perfectly mixed in the ISM - this ignores any form of dust clumping or its segregation compared to star forming sites. Secondly,  we assume all the energy resulting from SN events is instantaneously deposited in the ISM and contributes to sputtering for 30 Myr likely over-estimating the impact of this process. Thirdly, our model assumes equal amounts of the ISM gas to exist in a cold and dense versus a hot and diffuse phase, neglecting local density/temperature fluctuations. Incorporating a multi-phase ISM utilising clumping factors, could significantly impact our results by altering e.g. the efficiency of ISM dust growth. Finally, dust ejection is negligible up to $z \sim 13$, except for cases of highly increased star formation efficiency compared to the average population. Radiative SN feedback \citep{2023MNRAS.520.2445Z} could be more effective than the mechanical feedback prescription adopted in this work. Such enhancements are deferred to a future work.

\section*{Data Availability}
Data available on request to the corresponding author.


\section*{acknowledgments}
GN acknowledges financial support from the Eugenides Foundation in the form of a scholarship. GN is funded by the Cosmic Dawn Center (DAWN), in turn funded by the
Danish National Research Foundation under grant DNRF140. GN is also partially
funded by the European Union (ERC, HEAVYMETAL, 101071865). Views and
opinions expressed are however those of the authors only and do not necessarily
reflect those of the European Union or the European Research Council. Neither
the European Union nor the granting authority can be held responsible for them. PD acknowledges support from the NWO grant 016.VIDI.189.162 (``ODIN") and from the European Commission's and University of Groningen's CO-FUND Rosalind Franklin program. PD also warmly thanks the Institute for Advanced Study (IAS) Princeton, where a part of this work was carried out, for their generous hospitality and support through the Bershadsky Fund.
\bibliographystyle{mnras}
\bibliography{mybib}

\label{lastpage} 
\appendix


\section{The redshift evolution of the stellar mass function}

\begin{figure*}
\center{\includegraphics[scale=0.37]{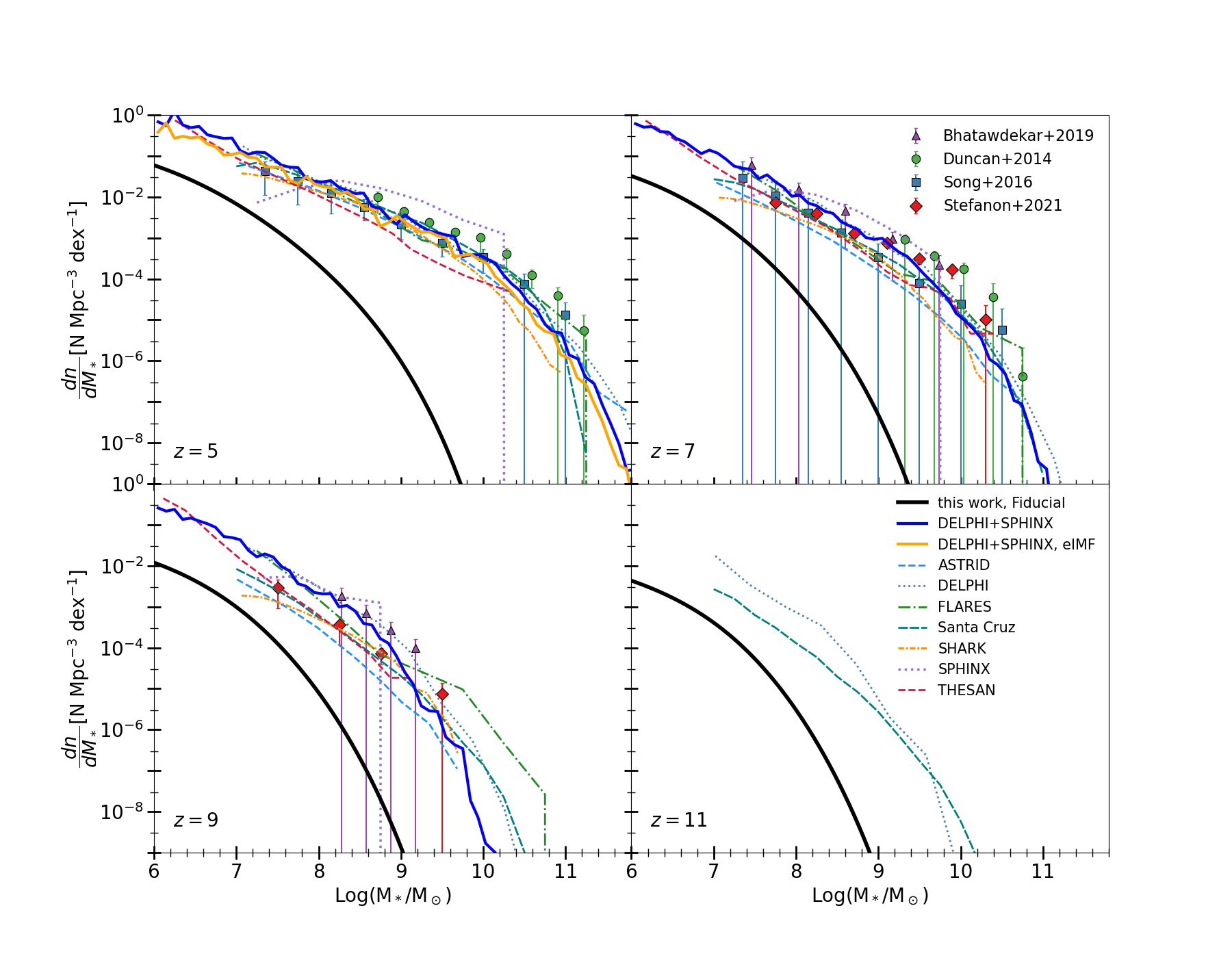}} 
\caption{The redshift evolution of the stellar mass function at $z \sim 5-11$. In all panels, the black line shows results from the {\it fiducial} model in this work. We show theoretical predictions from cosmological hydrodynamical
simulations including ASTRID (pale blue, dashed; \citealt{Bird2022}), and 
FLARES (green, dash-dotted; \citealt{Lovell2021}),
SPHINX (purple, dotted; \citealt{Katz2023}), and
THESAN (crimson, dashed; \citealt{kannan2021}), 
semi-analytical models DELPHI (blue, dotted;
\citealt{mauerhofer23dust}), Santa Cruz (teal, long-dashed; \citealt{Somerville2025}), and SHARK (orange, dash-dotted; \citealt{Lagos2024}). The solid blue and orange lines correspond to a coupling of the DELPHI and SPHINX models, for the fiducial case presented in \citet{Mauerhofer2025}, and an evolving IMF respectively. Observational data points are retrieved from \citet{duncan2014}, \citet{song2016}, \citet{bhatawdekar2019}, and \citet{stefanon2021}, as denoted in the legend.}
\label{fig_uvlf_with_models}
\end{figure*}

We compare the stellar mass function obtained from this work with a number of theoretical models and with observational data at $z \sim 5-11$. As shown in Fig. \ref{fig_uvlf_with_models}, the UV-LF calibrated analytic model from this work, is substantially different from that found in other works that range from using hydrodynamic simulations \citep{Lovell2021,kannan2021,Bird2022,Katz2023} to semi-analytic models \citep{mauerhofer23dust,Lagos2024,Somerville2025,Mauerhofer2025}, with the difference stemming from their distinct calibration strategies. In our model, the stellar mass hosted by a halo is calculated using the star formation efficiency $f_*(z)$, which is explicitly tuned to match the observed UV LF in the low luminosity (relatively dust-free) regime. In contrast, for example, \citet{Mauerhofer2025} import gas fractions, star formation efficiencies and feedback physics from the \code{sphinx-20} hydrodynamic simulation and evolve their galaxies inside the \code{delphi} semi-analytic framework, tracking their full stellar mass assembly history. This ``tracking" of the stellar mass assembly holds true for all of the other models noted here. Because UV light traces recent star formation, UV-derived masses preferentially trace young stellar populations, potentially missing out an older component dominating a galaxy’s mass. This can lead to the underestimation of a galaxy’s stellar mass by orders of magnitude, especially in bursty star forming systems \citep[see e.g.][]{topping_rebels}. That systematic offset in the stellar mass function therefore propagates directly into the dust mass function, as $\rm{M_{d}} \propto \rm{M_*}$ to first order, naturally explaining why our dust mass function sits below the predictions from the other models noted here. As a result, our dust mass function, though physical, yields a lower limit on the dust mass that a halo of a given mass can host across redshift. One way of matching both the UV LFs and stellar mass functions is to assume stellar populations become increasingly younger with increasing redshifts \citep{donnan2025}.

\end{document}